# Decentralization of Energy Systems with Blockchain: Bridging Top-down and Bottom-up Management of the Electricity Grid

Sakshi Mishra[1], Roohallah Khatami, Yu Christine Chen

Department of Electrical and Computer Engineering

The University of British Columbia, Vancouver, BC, Canada

---

[1]Corresponding author: sakshi.m@outlook.com

# Abstract

For more than a century, the grid has operated in a centralized top-down fashion. However, as distributed energy resources (DERs) penetration grows, the grid edge is increasingly infused with intelligent computing and communication capabilities. Thus, the bottom-up approach to grid operations inclined toward decentralizing energy systems will likely gain momentum alongside the existing centralized paradigm. Decentralization refers to transferring control and decision-making from a centralized entity (individual, organization, or group thereof) to a distributed network. It is not a new concept - in energy systems context or otherwise. In the energy systems context, however, the complexity of this multifaceted concept increases manifolds due to two major reasons - i) the nature of the commodity being traded (the electricity); and ii) the enormity of the traditional electricity sector's structure that builds, operates, and maintains this capital-intensive network.

In this work, we aim to highlight the need for and outline a credible path toward restructuring the current operational architecture of the electricity grid in view of the ongoing decentralization trends with an emphasis on peer-to-peer energy trading. We further introduce blockchain technology in the context of decentralized energy systems problems. We also suggest that blockchain is an effective technology for facilitating the synergistic operations of top-down and bottom-up approaches to grid management.

# 1 Background and Introduction

Recent academic research has focused on local decentralized energy trading. Researchers have tackled the solution to the question of efficient peer-to-peer (P2P) energy trading markets with various approaches. For example, a double-auction based game theoretic approach is employed by [1] to conduct P2P energy trading. Wang et al. [2], on the other hand, employ energy crowd-sourcing technique to enable blockchain based P2P energy markets. P2P energy trading within the context of virtual power plants is explored in [3]. Alashery et at. [4] proposed a quasi-ideal P2P trading framework using Blockchain. It has also been proposed and demonstrated that the management of decentralized energy systems can be facilitated using the blockchain technology [5] [6]. Prosumer preferences are also beginning to be considered in designing the P2P energy trading platforms [7]. The policy considerations for the use of Blockchain technology in P2P energy markets is also actively being studied [8].

There are also test-bed and prototype deployments around the world to demonstrate the feasibility of localized peer-to-peer energy trading [9]. Brooklyn microgrid being one of the first successful prototype, incorporating Blockchain technology for P2P energy trading, that's been operational for a few years [10]. This growing trend toward the bottom-up approach will challenge the institution of the traditional top-down architecture for grid operations. Easily anticipated are frictions that will arise amongst different actors along the electricity supply chain and in electricity markets (e.g., generation owners, large utilities, small DER owners, policy makers, etc.) as decentralization escalates.

Critical in a credible and practical path forward is to coordinate the coexistence of the two seemingly contradictory approaches of grid management, toward the overarching goal of reforming the grid of the future as a reliable, efficient, decarbonized, and economical infrastructure to meet the growing global energy demand. In the following sections we identify the value proposition of blockchain technology in enabling such synergies through secure yet accessible digitalization of energy data. We further include detailed description of *peer-to-peer (P2P) energy trading* as a use case for blockchain technology. We then shed light on how blockchain technology can help to create P2P trading platforms that work in harmony with existing distribution grid operations (top-down approach) while empowering the end users to participate in energy trades in localized settings (bottom-up approach). We conclude the article by identifying key future research directions at the nexus of P2P energy trading and blockchain technology for bringing the proposed viewpoint to practice.

# 2  Motivation and Perspective

The electricity sector has primarily been monopolistic dominated by vertically integrated utilities. Liberalization of electricity markets over the past several decades has opened doors *only* for owners of utility- scale generation plants to participate. From the *end user's perspective*, the electricity sector remains largely centralized: end users have very little (if at all) say in large-scale energy market dynamics and therefore they have no choice but to be "price-takers". The proliferation of DERs, reaching high penetrations, offers a beacon of hope: it empowers the end users and their communities in unprecedented ways toward self-sufficiency to the extent that going "off-grid" is starting to become a viable option. Depending on the jurisdiction, prosumers[2] may already have the option to sign contracts with *DER aggregators*–third-party *for-profit* entities serving as intermediaries between the prosumers and wholesale markets. Such contracts often lock prosumers into selling their electricity for a stipulated length of time (often a few years) and therefore does not necessarily maximize the value of energy that prosumers offer. If prosumers start mi- grating off-grid, it will cause a domino effect where fixed network costs will be shared by fewer remaining consumers, reducing the value for money offered to them by the network, thereby incentivizing them to also migrate off-grid. This is not desirable, at least in the coming few decades, for developed nations that have invested trillions of dollars over more than a century to erect and maintain a fully-fledged electricity grid network. Peer-to-peer (P2P) energy trading is an appealing option that becomes increasingly viable with greater DER penetration. It is a market apparatus that further spurs decentralization by empowering prosumers to have greater control on the use of energy generated by the DERs they own. Thus, P2P energy markets will play a significant role in propelling the bottom-up approach to energy systems operations.

The past decade has borne witness to significant [11]–[13] push back from utilities entrenched in the top-down architecture against DERs and community microgrids as these concepts challenge existing business models. In this point of view, we put forth the idea that the centralized (top-down) vs. decentralized (bottom-up) architectures for grid operations are not necessarily at fundamental odds.

> We propose that essential to a paradigm shift in the electricity sector's operating philosophy is to find synergies between top-down and bottom-up approaches of managing the generation, transmission, distribution, and end-use pipeline.

In practical terms, this proposition entails that the prosumers or the end users are no longer passive price-takers. Instead, they participate in the wholesale markets as price-makers.

At first glance, the idea of top-down and bottom-up approaches coexisting may seem paradoxical, and we may be tempted to label the bottom-up approach of infusing the grid edge with greater control coupled with DERs deployments as "disruptive". However, we must be mindful of the fact that technological disruptions in the electricity sector cannot and will not happen overnight given the conservative nature and sheer scale of this industry[3,4]. If the bottom-up operational approach leans toward augmenting the objectives the current top-down approach, rather than outright displacing the centralized system, it will more likely gain widespread adoption. The marriage of top-down and bottom-up architectures will require novel coordination mechanisms that can aggregate small-scale DERs toward system-wide objectives while fully preserving their owners' energy choices (i.e., when and how they would like to use the energy from their DERs). We argue that blockchain technology can serve as a medium to establish such dynamic prosumer participation platforms.

---

[2]Prosumers are "proactive consumers" who own on-site energy production and storage capabilities. They undertake a proactive behavior by managing their consumption, production through DERs and energy storage.

[3]In 2019, the electric power industry in the United States generated a revenue of about 401.7 billion U.S. dollars [14].

[4]According to the International Energy Agency, total investment in the U.S. energy sector was valued at $350 billion in 2018 (the second-largest in the world) [15].

# 3 Electric Grid Operations

The operational paradigm of today's electric grid can be classified as top-down and bottom-up approaches to grid management. The top-down architecture has existed for long (century old, centralized grid infrastructure) and the bottom-up architecture is beginning to take shape with increasing DER penetration acting as a harbinger of decentralization in electricity sector. We delineate on the nature of the top-down and bottom-up operational architectures in the following subsections.

## 3.1 Top-down Architecture

The price of the electricity and the market dynamics are determined by the major players (suppliers and wholesale buyers). The end users, who are becoming prosumers, do not have a voice in the wholesale markets. This approach works as follows: Large-scale suppliers (centralized generation owners, independent power producers with centralized plants, retail load serving entities, etc.) bid their energy into the markets that are operated by independent system operators (ISOs). The price of electricity is then set based on the prices that sellers and buyers bid into the market. Although there exists so-called *bilateral contracts* that bypass the wholesale market, it is out-of-reach for the end user as only major players have access to the information, tools, and resources to set up contracts.

## 3.2 Bottom-up Architecture

Small-scale generation (i.e., DERs) owners can make choices about how and when they buy (sell) their energy and from (to) whom. They are no longer passive "price-takers". This approach works as follows: Prosumers have access to a well-managed localized marketplace to trade their energy. There are also non-binding agreements with aggregators that can represent the prosumers in lucrative wholesale markets. The energy usage patterns of prosumers are not directly or indirectly controlled or modified by centralized utility in a top-down fashion (for example, with mechanisms like direct load control demand response programs [16]). Instead, prosumers themselves can manage their on-site generation resources, and potentially energy storage devices in such a way that their energy costs are minimized while not compromising with the usage preferences or patterns.

## 3.3 Confluence of Top-down and Bottom-up Architectures

The bottom-up approach will become commonplace with growing number and size of DER installations, at which point it will organically reach a sufficient level of influence to affect wholesale markets. The major challenge lies in creating a suitable ecosystem for the bottom-up approach to flourish *alongside* the existing top-down architecture. Key to such a setup are accessible choices that empower end users, which can be catalyzed with the following systems working in tandem with upstream markets:

- Establishing behind-the-meter energy management systems infused with grid-edge intelligence, for coordinating DERs on-site to meet end-user preferences.

- Building digital marketplaces for localized energy trading–the energy management systems in previous point must be interoperable with these digital platforms.

- Creating *autonomous* or *semi-autonomous* digital aggregators that can represent small-scale DERs in the wholesale markets–these digital aggregators must be interoperable for being able to exchange data with the localized marketplaces.

For the systems described above to materialize, not only are technological innovations crucial, but also policy, regulators, and government are needed to support their deployment in the electricity industry.

It is noteworthy that there is an existing mechanism called *Demand Response* which is a voluntary decrease in electrical consumption by end-use customers that is generally triggered by compromised grid reliability or high wholesale market prices. In exchange for conducting (and sometimes just committing) to curtail their load, customers are remunerated. Demand response, in limited ways, offers a way for consumers to actively participate in grid management.

# 4 Blockchain Technology and Applications in Energy Systems

## 4.1 The Technology

Blockchain is essentially a list of records stored in "blocks" of data where each block contains a header and points to the previous block, forming a chain of blocks. Trust and security are established in blockchain (a decentralized system) by having a process to i) validate, ii) verify, and iii) confirm transactions. The process starts with recording the transaction (which can be an exchange of any value such as currency, energy, data, etc.) in a distributed ledger of blocks. These blocks are designed to be tamper-proof as they are chained together through the process of hashing[5]. A new block can only be added to the chain after the peer nodes have reached a consensus on its authenticity, which is checked through a predefined consensus protocol.

Fundamentally, blockchain is a data structure that makes it possible to create a tamper-proof, distributed, peer-to-peer system of ledgers containing immutable, time-stamped and cryptographically connected blocks of data. In practice, this means that data can be written only once onto a ledger, which is then read-only for each user.

## 4.2 The Functionality

Blockchain is a cryptographically secured distributed ledger which has *write once, append only* system of storing information. It is distributed in nature and is completely or partially replicated on the peers nodes in the network. It is a fully decentralized system in its purest form[6]. The most utilized blockchain protocols, such as Ethereum networks, maintain and update their distributed ledgers in a decentralized manner, in stark contrast to traditional networks that rely on a trusted and centralized data repository. In structuring the network in this way, blockchain functions to remove the need for a trusted third party to handle and store transaction data. Instead, data are distributed so that every user has access to the same information at the same time.

## 4.3 The Value Proposition

Removing the need for trusted third party or intermediary is key to creating a decision-making environment in grid operations where all the needed information is available to all the market players (small-scale and large-scale owners and consumers alike). At the same time, it is important to acknowledge the importance of data security in the electricity sector because energy systems are part of the critical infrastructure of nations. Although the most utilized blockchain protocols to date, the Bitcoin and Ethereum networks, are *public blockchains*, it is technically feasible to design the architecture of blockchain platforms that do not expose all the data publicly yet retain its property of immutability. This will make it possible to share the data with all the players of the electricity market while maintaining its security.

---

[5]Hashing - hashing means taking an input string of any length and giving out an output of a fixed length. In the context of cryptocurrencies like bitcoin, the transactions are taken as input and run through a hashing algorithm (bitcoin uses SHA-256) which gives an output of a fixed length.

[6]By purest form, we mean a public blockchain. Blockchains can either be public, private, or consortium. Based on this type of categorization, they can either be partially or fully decentralized.

## 4.4 The Applications

Given the strong value proposition of blockchain technology in the electricity sector for *securely* storing and *immutably* sharing the energy generation and consumption data, its applications are wide ranging. It has the potential to benefit the existing operations of various actors within both top-down and bottom-up approaches to grid management.

### 4.4.1 Applications in the top-down sphere:

- Operation of utilities can be made more effective using blockchains in multiple aspects including automated billing, interactions with smart DERs, etc.
- Many tasks related to wholesale energy trading and supply involving ISOs can be automated using smart contracts.
- The time-frame of settling power imbalances can be vastly reduced by using blockchain platform instead of manual calculations.
- Smart contracts could potentially simplify and speed up switching of energy suppliers by the consumers/end-users. Enhanced mobility of the consumers, in regards to choosing their energy supplier in the market, could further spur competition in wholesale and retail markets. This can potentially reduce energy tariffs.
- Immutable records and transparent (within the bounds of organization) critical infrastructure information storage mechanisms can significantly improve auditing and regulatory compliance process for utilities.

### 4.4.2 Applications in the bottom-up sphere:

- Blockchains facilitate digital P2P transactions, and they can potentially enable machine-to-machine (M2M) communication and data exchanges between smart devices.
- Sharing of infrastructure and clean energy resources at the distribution grid level: shared electric vehicle charging stations and community storage. Blockchain based digital platforms, offered as mobile application, can help facilitate effective utilization of such shared infrastructure.
- Perhaps the most promising application of blockchain technology in energy systems is the establishment of P2P energy markets. They hold tremendous potential for empowering prosumers to make their energy choice, satisfy their preferences, and yet do so in a cost effective manner.

Lastly, blockchain does not only have top-down and bottom-up architecture specific applications (as listed above); it can, perhaps more importantly, serve as a digital bridge between top-down and bottom-up operational architectures to help them run in parallel, synergistically. This is because, at its core, blockchain technology offers a secure and tamper-proof way to exchange value (value can be currency or energy systems data in our use-case) digitally without the need of an intermediary. Secure and near-real-time exchange of energy data [7] is the key to connecting the cyber and market layers of top-down and bottom-up grid management systems.

---

[7] Energy data constitutes multiple quantities such as:

1. Measured energy production and consumption
2. Measured currents and voltages
3. Predicted energy production and consumption
4. Electrified transportation: vehicle location, planned routes, and charging times/schedules
5. Expected participation in demand response

# 5 Peer-to-Peer Energy Trading Using Blockchain

P2P energy markets capacitate the prosumers to trade electricity in the absence of intermediaries at their agreed price, by establishing a platform to transact with each other. It is an appealing option that becomes increasingly viable with greater DER penetration. This market apparatus further spurs decentralization by empowering prosumers to have greater control over the use of energy generated by the DERs they own. Thus, P2P energy markets will play a significant role in propelling the bottom-up approach to energy systems operations. Moreover, when designed appropriately, P2P energy markets not only empower the prosumers but could potentially offer many benefits to the central grid such as aid with congestion management and providing ancillary services.

Although capable of functioning as a separate mechanism, peer-to-peer energy markets are largely going to operate amongst prosumers who are connected to the utility grid. Therefore, when the P2P platform organizes trades amongst participants, it will need to interact with the system operator and the retail electricity market because of several reasons that include: i) power that flows between the participants will affect the local distribution network; ii) the local distribution network needs to be operated, maintained, and remunerated accordingly (i.e., the fixed cost component of the electricity bill that accounts for the asset cost and costs of maintaining a capital intensive power distribution system); iii) the excess generation that spills out and the net demand needing energy from distribution grid (in case the generation is the P2P market is not enough to serve the demand) from the P2P market will be transacted with the upstream network. A secure and fair P2P energy trading platform is the epitome of the bottom-up approach to grid operations, but it cannot operate independently of the existing distribution grid, as explained above. This highlights the central argument of this point-of-view article which is – finding ways to make *top-down* and *bottom-up* approaches to grid operations coexist in a systematic and holistic manner.

Blockchain has the potential to serve as the technical medium that can offer the needed digital platforms for P2P energy markets to interact effectively with the central grid and its wholesale markets. 1 shows the overall P2P energy trading market architecture. These markets operate on a complex interplay of the three layers depicted in the diagram. There is a near-real-time flow of information between the three layers (vertical) as well as the participants (horizontal).

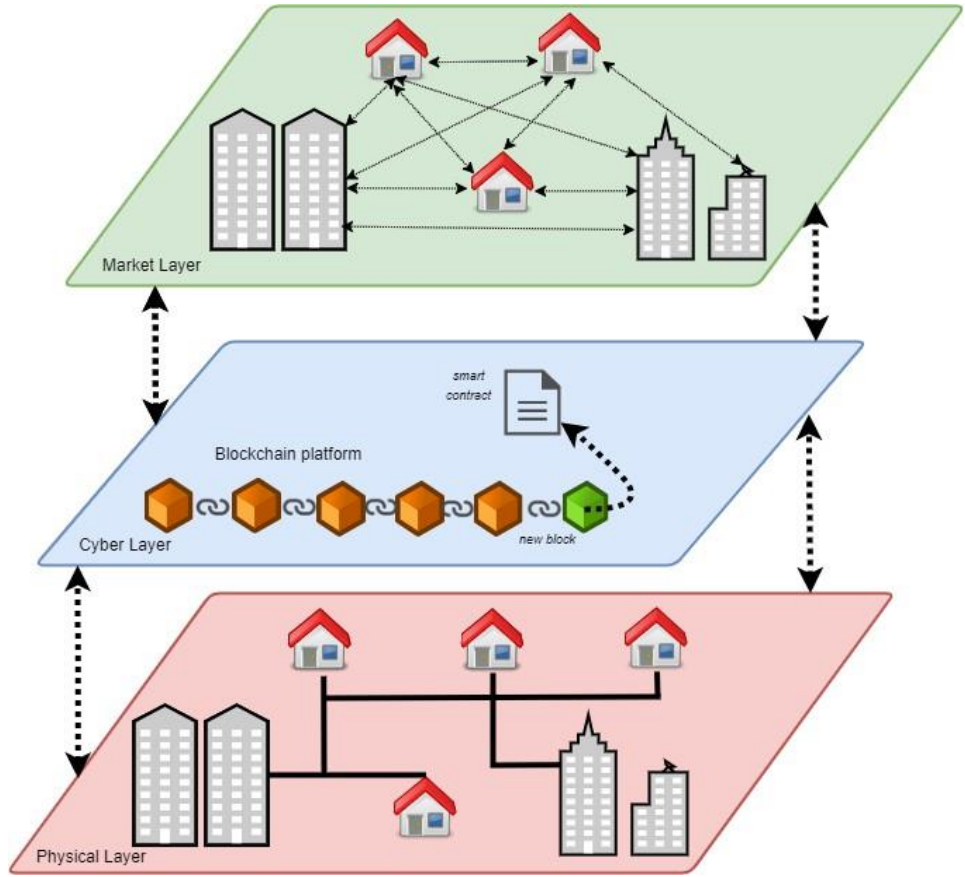

Figure 1:
P2P energy market operations constitute of three intertwined layers: market, cyber (blockchain), and physical.

# 6 Conclusion and Outlook

The proposed viewpoint represents an important interdisciplinary challenge that requires input from various traditionally distinct fields of study. Certainly, engineering advances in the application of blockchain technology in energy sector will be vital in enabling the paradigm shift in grid operations. This requires the coordinated efforts of researchers from fields like computer science, networking and communications, as well as electrical and power systems. Moreover, research in disciplines like policy and economics coupled with regulatory changes are needed to approach the transition between top-down and bottom-up architectures in a holistic and systematic manner.

There are ongoing efforts to integrate the DERs more effectively in the grid. The recent FERC ruling (Order No. 2222) [17] has, opportunely, enabled DERs to participate in the wholesale markets. This rule enables DERs to participate alongside traditional resources in the regional organized wholesale markets through aggregations, opening U.S. organized wholesale markets to new sources of energy and grid services. But there are a lot more facets of energy systems' current operational paradigm that need to be enhanced to truly integrate the top-down and bottom-up approaches to grid management. Here we identify three key future research directions on the nexus of P2P energy trading and blockchain technologies that are essential for bringing the vision of large-scale decentralized energy systems to reality.

- Fully autonomous and decentralized peer-to-peer energy markets: The market designs presented in the recent literature as well as pilot projects lack the property of being decentralized to the extent that the distribution system operator (DSO) are no longer needed to schedule the dispatch of the assets. As long as DSO are scheduling the dispatch of resources in P2P markets, these markets are operating under a *central authority*. This creates the problems such as *single point of failure in the network* and vulnerability towards cyber attacks disrupting the operations by *hacking the central dispatch system*.

Therefore, there is a room for lot of innovation in this direction such that these markets are executed on blockchain based platforms hosting the smart contracts and automated dispatch algorithms instead of manual operators making decisions by running applications on their computers.

- Integration of self-sovereign identity-based electric vehicles: For electric vehicles (EVs) to participate in P2P Energy Markets, it is important that there are mechanisms for them to interface with the market platform *on the move*. EVs shouldn't be constrained to be at a specific location (typically behind the customer meter of the home owner) to sell their energy to the grid. Instead, having access to the blockchain based energy trading platform via self-sovereign identity will empower EV owners to trade energy (buy or sell) without strict geographical constraints.

- Interoperability: Blockchain interoperability across different blockchains is required for moving energy data for different purposes (forecasting, scheduling, enforcing blockchain based contracts, etc.). For the localized P2P energy trading application, interoperability is a crucial functionality for ensuring sustainable growth and expansion of such localized energy trading platform. An added benefit of enabling interoperability is the flexibility to incorporate different blockchain technologies within the same market, thereby future proofing our platform to likely expansions in the fast evolving field of distributed ledger technologies. Therefore, solving the blockchain interoperability conundrum can have a significant impact on enabling P2P energy markets.

# References

[1] H. T. Doan, J. Cho, and D. Kim, "Peer-to-peer energy trading in smart grid through blockchain: A double auction-based game theoretic approach," *IEEE Access*, vol. 9, pp. 49 206–49 218, 2021. DOI: 10.1109/ACCESS.2021.3068730.

[2] S. Wang, A. F. Taha, J. Wang, K. Kvaternik, and A. Hahn, "Energy crowdsourcing and peer-to-peer energy trading in blockchain-enabled smart grids," *IEEE Transactions on Systems, Man, and Cybernetics: Systems*, vol. 49, no. 8, pp. 1612–1623, 2019. DOI: 10.1109/TSMC.2019.2916565.

[3] S. Seven, G. Yao, A. Soran, A. Onen, and S. M. Muyeen, "Peer-to-peer energy trading in virtual power plant based on blockchain smart contracts," *IEEE Access*, vol. 8, pp. 175 713–175 726, 2020. DOI: 10.1109/ACCESS.2020.3026180.

[4] M. K. AlAshery, Z. Yi, D. Shi, *et al.*, "A blockchain-enabled multi-settlement quasi-ideal peer-to-peer trading framework," *IEEE Transactions on Smart Grid*, vol. 12, no. 1, pp. 885–896, 2021. DOI: 10.1109/TSG.2020.3022601.

[5] L. Thomas, Y. Zhou, C. Long, J. Wu, and N. Jenkins, "A general form of smart contract for decentralized energy systems management," *Nature Energy*, vol. 4, no. 2, pp. 140–149, Feb. 2019, ISSN: 2058-7546. DOI: 10.1038/s41560-018-0317-7. [Online]. Available: https://doi.org/10.1038/s41560-018-0317-7.

[6] D. Han, C. Zhang, J. Ping, and Z. Yan, "Smart contract architecture for decentralized energy trading and management based on blockchains," *Energy*, vol. 199, p. 117 417, 2020, ISSN: 0360-5442. DOI: https://doi.org/10.1016/j.energy.2020.117417. [Online]. Available: https://www.sciencedirect.com/science/article/pii/S0360544220305247.

[7] T. Morstyn and M. D. McCulloch, "Multiclass energy management for peer-to-peer energy trading driven by prosumer preferences," *IEEE Transactions on Power Systems*, vol. 34, no. 5, pp. 4005–4014, 2019. DOI: 10.1109/TPWRS.2018.2834472.

[8] U. Cali and O. Çakir, "Energy policy instruments for distributed ledger technology empowered peer-to-peer local energy markets," *IEEE Access*, vol. 7, pp. 82 888–82 900, 2019. DOI: 10.1109/ACCESS.2019.2923906.

[9] M. Andoni, V. Robu, D. Flynn, *et al.*, "Blockchain technology in the energy sector: A systematic review of challenges and opportunities," *Renewable and Sustainable Energy Reviews*, vol. 100, pp. 143–174, 2019, ISSN: 1364-0321. DOI: https://doi.org/10.1016/j.rser.2018.10.014. [Online]. Available: https://www.sciencedirect.com/science/article/pii/S1364032118307184.

[10] E. Mengelkamp, J. Gärttner, K. Rock, S. Kessler, L. Orsini, and C. Weinhardt, "Designing microgrid energy markets: A case study: The brooklyn microgrid," *Applied Energy*, vol. 210, pp. 870–880, 2018, ISSN: 0306-2619. DOI: https://doi.org/10.1016/j.apenergy.2017.06.054. [Online]. Available: https://www.sciencedirect.com/science/article/pii/S030626191730805X.

[11] United States Court of Appeals for the district of Columbia circuit. "National association of regulatory utility commissioners, petitioner v. federal energy regulatory commission, respondent." Accessed: 2021-07-08. (Jul. 2020), [Online]. Available: https://www.cadc.uscourts.gov/internet/opinions.nsf/E12B1903B0477E21852%5C$file/19-1142-1851001.pdf.

[12] Gavin Bade. "Miso states push for authority over der in electricity markets at ferc meeting." Accessed: 2021-07-08. (Apr. 2018), [Online]. Available: https://www.utilitydive.com/news/miso-states-push-for-authority-over-der-in-electricity-markets-at-ferc-meet/521089/.

[13] Jim Matheson. "Will ferc trample state and local authorities in der rulemaking?" Accessed: 2021-07-08. (Apr. 2019), [Online]. Available: https://www.utilitydive.com/news/will-ferc-trample-state-and-local-authorities-in-der-rulemaking/552069/.

[14] Bruna Alves. "Revenue of the electric power industry in the united states from 1970 to 2019." Accessed: 2021-07-08. (Jul. 2021), [Online]. Available: https://www.statista.com/statistics/190548/revenue-of-the-us-electric-power-industry-since-1970/.

[15] SELECT USA. "Energy industry spotlight - the energy industry in the united states." Accessed: 2021-07-08. (), [Online]. Available: https://www.selectusa.gov/energy-industry-united-states.

[16] D. Li, W.-Y. Chiu, and H. Sun, "Chapter 7 - demand side management in microgrid control systems," in *Microgrid*, M. S. Mahmoud, Ed., Butterworth-Heinemann, 2017, pp. 203–230, ISBN: 978-0-08-101753-1. DOI: https://doi.org/10.1016/B978-0-08-101753-1.00007-3. [Online]. Available: https://www.sciencedirect.com/science/article/pii/B9780081017531000073.

[17] FERC. "Ferc order no. 2222: A new day for distributed energy resources." (2020), [Online]. Available: https://www.ferc.gov/media/ferc-order-no-2222-fact-sheet.